\documentstyle[11pt]{article}

\newcommand{\B}{{\cal B}}

\begin{document}
\begin{center}
{\LARGE{\bf Energy and Momentum Distributions of Kantowski and Sachs Space-time }}\\[2em]
\large{\bf{Ragab M. Gad\footnote{Email Address: ragab2gad@hotmail.com} and A. Fouad}}\\
\normalsize {Mathematics Department, Faculty of Science,}\\
\normalsize  {Minia University, 61915 El-Minia,  EGYPT.}
\end{center}

\begin{abstract}
We use the Einstein, Bergmann-Thomson, Landau-Lifshitz and
Papapetrou energy-momentum complexes to calculate the energy and
momentum distributions of Kantowski and Sachs space-time. We show
that the Einstein and Bergmann-Thomson definitions  furnish a
consistent result for the energy distribution, but the definition
of Landau-Lifshitz do not agree with them.  We show that a
signature switch should affect about everything including energy
distribution in the case of Einstein and Papapetrou prescriptions
but not in Bergmann-Thomson and Landau-Lifshitz prescriptions.

\end{abstract}

\setcounter{equation}{0}
\section{Introduction}
One of the most interesting and intricate problems still unsolved
since the outset of general relativity is the energy-momentum
localization. Einstein himself proposed the first energy-momentum
complex in an attempt to define the local distribution of energy and
momentum \cite{E}. After this attempt, a plethora of different
energy-momentum complexes were proposed, including formulations by
Tolman \cite{T}, Papapetrou \cite{P}, M{\o}ller \cite{E1}, Landau
and Lifshitz \cite{LL}, Weinberg \cite{W}, Bergmann-Thomson \cite{B}
and others. This approach was abandoned for a long time due to
severe criticism for a number of reasons.
\par
Recently, Virbhadra  re-opened  the subject of energy-momentum
complexes \cite{Virb}. He pointed out that though these complexes
are non-tensors, they yield reasonable and consistent results for
a given space-time. Aguirregabiria et al \cite{AV} found that for
any metric of the Kerr-Schild class, several different definitions
of the energy-momentum complex yield precisely the same
results.Virbhadra \cite{V99} investigated whether or not these
energy momentum complexes lead to the same results for the most
general non-static spherically symmetric metric and found that
they disagree. He noted that the energy-momentum complexes of
Landau and Lifshitz, Papapetrou and Weinberg give the same results
as in the Einstein definition if the calculations are performed in
Kerr-Schild Cartesian coordinates. However, these energy-momentum
complexes disagree if computations are done in "Schwarzschild
Cartesian" coordinates\footnote{ Schwarzschild metric in
``Schwarzschild Cartesian coordinates'' is obtained by
transforming this metric (in usual Schwarzschild coordinates $\{t,
r, \theta, \phi\}$)  to $\{t,x,y,z\}$ using $ x = r \sin\theta
\cos\phi, y = r \sin\theta \sin\phi, z = r \cos\theta $.}. In a
detail study of the question, Xulu \cite{Xulu1} has confirmed this
suggestion. He obtained the energy distribution for the most
general non-static spherically symmetric using M{\o}ller's
definition and found different results in general from those
obtained using Einstein's definition. These results agree for the
Schwarzschild, Vaidya and Janis-Newmann-Winicour space-times, but
disagree for the Reissner-Nordstr\"{o}m space-time. Many authors
had similarly successfully applied the aforementioned
energy-momentum complexes to various black hole configurations
\cite{Many}.
\par
It has been remained a controversial problem whether or not energy
and momentum are localizable. There are different opinions on this
subject. Contradicting the viewpoint of Misner et al. \cite{MTW}
that the energy is localizable only for spherical systems,
Cooperstock and Sarracino \cite{CS} argued that if the energy
localization is meaningful for spherical systems then it is
meaningful for all systems. Bondi \cite{Bo} expressed that a
non-localizable form of energy is inadmissible in relativity and
its location can in principle be found. These contradictory
viewpoints bear significantly on the study of gravitational waves.
It is an interesting question whether or not gravitational waves
have energy and momentum content. In a series of papers,
Cooperstock \cite{COO} hypothesized that in a curved space-time
energy and momentum are confined  to the region of non-vanishing
energy-momentum tensor $T^{a}_{b}$  and consequently the
gravitational waves are not carriers of energy and momentum in
vacuum space-times. This hypothesis has neither been proved nor
disproved. There are many results supporting this hypothesis (see
for example, \cite{Xulu,Gad2}).
\par
In this paper we evaluate the energy and momentum distributions of
the Kantowski and Sachs space-time, using Einstein,
Bergmann-Thomson, Landau-Lifshitz and Papapetrou energy-momentum
complexes.
\par
Through this paper we use $G = 1$ and $c = 1$ units and follow the
convention that Latin indices take value from 0 to 3 and Greek
indices take value from 1 to 3.
\section{Kantowski and Sachs Space-time}
The  standard representation of Kantowski and Sachs space-times
are given by \cite{KS}
\begin{equation} \label{K-S}
dS^2= dt^2 - A^2(t) dr^2 - B^2(t)\big(d^2\theta + \sin^2\theta
d^2\phi\big),
\end{equation}
where the functions $A(t)$ and $B(t)$ are to be determined from
the field equations.\\
The solutions of the Einstein field equations for the above metric
were considered with dust source \cite{KS}, but they were
generalized (in fact earlier) to the general perfect fluid source
\cite{KC}.
\par
From the geometrically point of view, this line element admits a
four parameter continuous group of isometries which acts on
space-like hypersurface, and has no three parameter subgroup that
would be simply transitive on the orbits  (for more detailed
description see Kantowski and Sachs \cite{KS} and Collins \cite{C77}).\\
From the physical point of view, the metric (\ref{K-S})
automatically defines an energy-momentum tensor of a fluid with
anisotropic pressure, and the coordinates of (\ref{K-S}) are
comoving. The rotation and acceleration are zero, but if the
source is to be a perfect fluid, then the shear is necessarily
non-zero.
\par
It is well known that if the calculations are performed in
quasi-Cartesian coordinates,
all the energy-momentum complexes give meaningful results.\\
According to the following transformations
$$
r = \sqrt{x^2 + y^2+ z^2} , \qquad \phi = \arctan(\frac{y}{x}),
$$
the line element (\ref{K-S}) written in terms of  quasi-Cartesian
coordinates reads:
\begin{equation} \label{2}
dS^2 = dt^2 + \frac{B^2}{r^2}\big( dx^2 + dy^2+dz^2 \big) -
\frac{1}{r^2}\Big(A^2-\frac{B^2}{r^2}\Big)\big(xdx +ydy+zdz\big)^2.
\end{equation}

\par
 For the above metric the determinant of the metric tensor
and the contravariant components of the tensor are given,
respectively, as follows
\begin{equation}\label{3}
\begin{array}{ccc}
det(g) & = - \frac{A^2B^4}{r^4},\\
g^{00} & = 1, \\
g^{11} & = x^2\Big(\frac{1}{B^2}-\frac{1}{r^2A^2}\Big)-\frac{r^2}{B^2},\\
g^{12} & =  xy\Big(\frac{1}{B^2} - \frac{1}{r^2A^2}\Big),\\
g^{13} & =  xz\Big(\frac{1}{B^2}-\frac{1}{r^2A^2}\Big),\\
g^{22} & = y^2\Big(\frac{1}{B^2}-\frac{1}{r^2A^2}\Big)-\frac{r^2}{B^2},\\
g^{23} & =  yz\Big(\frac{1}{B^2}-\frac{1}{r^2A^2}\Big),\\
g^{33} & =
z^2\Big(\frac{1}{B^2}-\frac{1}{r^2A^2}\Big)-\frac{r^2}{B^2}.
\end{array}
\end{equation}

\section{Einstein's Energy-momentum Complex}
The energy-momentum complex as defined
by Einstein \cite{E} is given by
\begin{equation} \label{3.1}
\theta^{k}_{i} = \frac{1}{16\pi}H^{kl}_{\, \,\, i,l},
\end{equation}
where the Einstein's superpotential $H^{kl}_{\, \,\, i}$ is of the
form
\begin{equation} \label{3.2}
H^{kl}_{\, \,\, i} = - H^{lk}_{\, \,\, i} = \frac{g_{in}}{\sqrt{-
g}} \big[ - g\big( g^{kn}g^{lm} - g^{ln}g^{km}\big)\big]_{,m}.
\end{equation}
$\theta^{0}_{0}$ and $\theta^{0}_{\alpha}$ are the energy and
momentum density components, respectively.\\
The Einstein energy-momentum satisfies the local conservation law
$$
\frac{\partial\theta^{k}_{i}}{\partial x^k}=0.
$$
The energy and momentum in the Einstein's prescription are given
by
\begin{equation}
P_{i}=\int\int\int \theta^{0}_{i}dx^1 dx^2 dx^3.
\end{equation}
Using the Gauss theorem we obtain
\begin{equation}\label{6}
P_{i}=\frac{1}{16\pi}\int\int H^{0\alpha}_{i}n_{\alpha}ds,
\end{equation}
where $n_{\alpha}=(\frac{x}{r},\frac{y}{r},\frac{z}{r})$ are the
components of a normal vector over an infinitesimal surface
element $ds=r^2\sin\theta d\theta d\phi$.
\par
The required  non zero components of $H^{kl}_{\, \,\, i}$ for the
line element (\ref{K-S}) are given by
\begin{equation}\label{7}
\begin{array}{ccc}
H^{01}_{\, \,\, 0} = \frac{2x}{Ar^4}\Big(A^2r^2+B^2\Big),\\
H^{02}_{\, \,\, 0} =\frac{2y}{Ar^4}\Big(A^2r^2+B^2\Big),\\
H^{03}_{\, \,\, 0} =\frac{2z}{Ar^4}\Big(A^2r^2+B^2\Big),\\
H^{01}_{\, \,\, 1} =\frac{2B}{r^2}\Big(\frac{x^2}{r^2}(\dot{A} B-
A\dot{B})-(\dot{A} B+A\dot{B})\Big),\\
H^{01}_{\, \,\, 2} = H^{02}_{\, \,\, 1} =\frac{2xyB}{r^4}(\dot{A}
B-
A\dot{B}),\\
H^{01}_{\, \,\, 3} = H^{03}_{\, \,\, 1} =\frac{2xzB}{r^4}(\dot{A}
B-
A\dot{B}),\\
 H^{02}_{\, \,\, 2} =\frac{2B}{r^2}\Big(\frac{y^2}{r^2}(\dot{A} B-
A\dot{B})-(\dot{A} B+A\dot{B})\Big),\\
H^{02}_{\, \,\, 3} = H^{03}_{\, \,\, 2} =\frac{2yzB}{r^4}(\dot{A}
B-A\dot{B}),\\
H^{03}_{\, \,\, 3} = \frac{2B}{r^2}\Big(\frac{z^2}{r^2}(\dot{A} B-
A\dot{B})-(\dot{A} B+A\dot{B})\Big).
\end{array}
\end{equation}
 Using the components (\ref{7}) we
obtain the components of energy and momentum densities in the form
\begin{equation}\label{8}
\begin{array}{ccc}
 \theta^{0}_{0}& =\frac{1}{8\pi Ar^4}(A^2r^2-B^2),\\
\theta^{0}_{1}& =\frac{Bx}{4\pi r^4}(\dot{A} B+A\dot{B}),\\
\theta^{0}_{2}& =\frac{By}{4\pi r^4}(\dot{A} B+A\dot{B}),\\
\theta^{0}_{3}& =\frac{Bz}{4\pi r^4}(\dot{A} B+A\dot{B}).
\end{array}
\end{equation}
Using equations (\ref{7}) in equation (\ref{6}),  the energy and
momentum distributions are the following
$$
E_{Ein}=P_{0}= \frac{1}{2Ar}\big(A^2r^2 + B^2\big),
$$
$$
P_{1}=P_{2}=P_{3}=0.
$$
We notice that if the signature of the space-time under study is
changed from +2 to -2, we find that the values of energy and
momentum densities as well as the energy distribution are changed
from positive to negative.

\setcounter{equation}{0}
\section{\bf{The Energy-Momentum Complex of Bergmann-Thomson}}
The Bergmann-Thomson  energy-momentum complex \cite{B} is given by
\begin{equation}\label{B.1}
{\bf{B}}^{ik} = \frac{1}{16\pi}\big[g^{il}\B^{km}_{l}\big]_{,m},
\end{equation}
where
$$
\B^{km}_{l} =
\frac{g_{ln}}{\sqrt{-g}}\Big[-g\Big(g^{kn}g^{mp}-g^{mn}g^{kp}\Big)\Big]_{,p}.
$$
$B^{00}$ and $B^{0\alpha}$ are the energy and momentum density
components.\\
The energy and momentum are given by
\begin{equation}
P^i=\int\int\int B^{i0}dx^1dx^2dx^3.
\end{equation}
Using the Gauss theorem we have
\begin{equation} \label{epB}
P^i=\frac{1}{16\pi}\int\int\B^{i0\alpha}n_{\alpha}dS.
\end{equation}

In order to calculate the energy and momentum distributions for
the space-time under consideration, using Bergmann-Thomson
energy-momentum complex, we require the following non-vanishing
components of $\B^{km}_{l}$
\begin{equation}\label{B}
\begin{array}{ccc}
\B^{01}_{\, \,\, 0} = \frac{2x}{Ar^4}\Big(A^2r^2+B^2\Big),\\
\B^{02}_{\, \,\, 0} =\frac{2y}{Ar^4}\Big(A^2r^2+B^2\Big),\\
\B^{03}_{\, \,\, 0} =\frac{2z}{Ar^4}\Big(A^2r^2+B^2\Big),\\
\B^{01}_{\, \,\, 1} =\frac{2B}{r^2}\Big(\frac{x^2}{r^2}(\dot{A} B-
A\dot{B})-(\dot{A} B+A\dot{B})\Big),\\
\B^{01}_{\, \,\, 2} = \B^{02}_{\, \,\, 1}
=\frac{2xyB}{r^4}(\dot{A} B-
A\dot{B}),\\
\B^{01}_{\, \,\, 3} = \B^{03}_{\, \,\, 1}
=\frac{2xzB}{r^4}(\dot{A} B-
A\dot{B}),\\
 \B^{02}_{\, \,\, 2} =\frac{2B}{r^2}\Big(\frac{y^2}{r^2}(\dot{A} B-
A\dot{B})-(\dot{A} B+A\dot{B})\Big),\\
\B^{02}_{\, \,\, 3} = \B^{03}_{\, \,\, 2}
=\frac{2yzB}{r^4}(\dot{A}
B-A\dot{B}),\\
\B^{03}_{\, \,\, 3} = \frac{2B}{r^2}\Big(\frac{z^2}{r^2}(\dot{A}
B- A\dot{B})-(\dot{A} B+A\dot{B})\Big).
\end{array}
\end{equation}

 Using the components (\ref{B})  in (\ref{B.1}),  the components of
energy and momentum densities are as follows
\begin{equation}\label{14}
\begin{array}{ccc}
{\bf{B}}^{00}& =\frac{1}{8\pi Ar^4}(A^2r^2-B^2),\\
{\bf{B}}^{01}& =-\frac{x}{8\pi r^4}\Big(\dot{A} r^2+\frac{B}{A^2}(2A\dot{B} -\dot{A} B\Big),\\
{\bf{B}}^{02}& =-\frac{y}{8\pi r^4}\Big(\dot{A} r^2+\frac{B}{A^2}(2A\dot{B} -\dot{A} B\Big),\\
{\bf{B}}^{03}& =-\frac{z}{8\pi r^4}\Big(\dot{A}
r^2+\frac{B}{A^2}(2A\dot{B} -\dot{A} B\Big).
\end{array}
\end{equation}
Using equations (\ref{B}) in equation (\ref{epB}), we obtain the
energy and  momentum distributions in the following form
$$
E_{Berg}=P^{0}= \frac{1}{2Ar}\big(A^2r^2 + B^2\big),
$$
$$
P_{1}=P_{2}=P_{3}=0.
$$
The above energy density and energy distribution are agreement
with that obtained before, using Einstein's energy-momentum
complex.
 In the case of Bergmann's energy-momentum complex, we notice that
 a signature  switch do not affect about everything including
energy distribution. Consequently, the Einstein and
Bergmann-Thomson prescription do not give the same results when
the signature of the space-time under study is -2.

\setcounter{equation}{0}
\section{\bf{Landau-Lifshitz's Energy-momentum Complex}}
Landau-Lifshitz's energy-momentum complex \cite{LL} is given by
\begin{equation}\label{5.1}
L^{ij} = \frac{1}{16\pi}S^{ikjl}_{\quad ,kl},
\end{equation}
where
\begin{equation}\label{5.2}
S^{ikjl} = -g\big( g^{ij}g^{kl} - g^{il}g^{kj}\big).
\end{equation}
$L^{ij}$ is symmetric in its indices, $L^{00}$ is the energy
density and $L^{0\alpha}$ are the momentum (energy current)
density components. $S^{ikjl}$ has the symmetries of the Riemann
curvature tensor.\\
The energy and momentum are given by
\begin{equation}
P^i=\int\int\int L^{i0}dx^1dx^2dx^3.
\end{equation}
Using the Gauss theorem we have
\begin{equation} \label{emLL}
P^i=\frac{1}{16\pi}\int\int S^{i0\alpha}n_{\alpha}dS.
\end{equation}

\par
The required non-vanishing components of $S^{ikjl}$ are
\begin{equation}\label{3.5}
\begin{array}{ccc}
S^{0101} & = B^2\Big[\frac{x^2}{r^6}(A^2r^2-B^2)-\frac{A^2}{r^2}\Big],\\
S^{0102} & = \frac{xy}{r^6}B^2(A^2r^2-B^2),\\
S^{0103} & = \frac{xz}{r^6}B^2(A^2r^2-B^2),\\
S^{0202} & = B^2\Big[\frac{y^2}{r^6}(A^2r^2-B^2)-\frac{A^2}{r^2}\Big],\\
S^{0203} & = \frac{yz}{r^6}B^2(A^2r^2-B^2),\\
S^{0303} & =
B^2\Big[\frac{z^2}{r^6}(A^2r^2-B^2)-\frac{A^2}{r^2}\Big].
\end{array}
\end{equation}
Using these components  in equation (\ref{5.1}), we obtained  the
energy  and momentum densities are
\begin{equation}
\begin{array}{ccc}
 L^{00}& =-\frac{B^2}{8\pi r^6}\big[A^2r^2+3B^2\big],\\
 L^{10}& =-\frac{xB}{4\pi r^6}\big[ A(A\dot{B} +\dot{A} B)r^2 +
 2B^2\dot{B}\big],\\
L^{20}& =-\frac{yB}{4\pi r^6}\big[ A(A\dot{B} +\dot{A} B)r^2 +
 2B^2\dot{B}\big],\\
 L^{30}& =-\frac{zB}{4\pi r^6}\big[ A(A\dot{B} +\dot{A} B)r^2 +
 2B^2\dot{B}\big].
\end{array}
\end{equation}
Using equations (\ref{5.1}) in equation (\ref{emLL}), we obtain
the energy and momentum distributions
\begin{equation}
E_{LL}=\frac{B^2}{2r^3}(A^2r^2+B^2).
\end{equation}
\begin{equation}
P^1_{LL}=P^{2}_{LL}=P^{3}_{LL}=0.
\end{equation}
The above results do not agree with the results obtained before,
using Einstein and Bergmann-Thomson energy-momentum complexes. A
signature switch does not affect about everything including energy
distribution.

\setcounter{equation}{0}
\section{\bf{Papapetrou's Energy-momentum Complex}} The symmetric
energy-momentum complex of Papapetrou \cite{P} is given by
\begin{equation}\label{P1}
\Omega^{ij} = \frac{1}{16\pi} \Upsilon^{ijkl}_{\quad ,kl},
\end{equation}
where
\begin{equation}\label{}
\Upsilon^{ijkl} = \sqrt{- g}\big( g^{ij}\eta^{kl} -
g^{ik}\eta^{jl} + g^{kl}\eta^{ij} - g^{jl}\eta^{ik}\big),
\end{equation}
and $\eta^{ik}$ is the Minkowski metric with signature $+2$.\\
$\Omega^{00}$ and $\Omega^{\alpha 0}$ are the energy and momentum
density components.
\\
The energy and momentum, using the Papapetrou prescription are
given by
\begin{equation}
P^i =\int\int\int\Omega^{i0}dx^1dx^2dx^3.
\end{equation}
Using the Gauss theorem we obtain
\begin{equation} \label{emP}
P^i =\frac{1}{16\pi}\int\int \Upsilon^{i0\alpha
l}_{\,\,\,\,\,\,,l} n_{\alpha} dS.
\end{equation}

 The non-vanishing components of
$\Upsilon^{ijkl}$  are as follows
\begin{equation}\label{P}
\begin{array}{ccc}
\Upsilon^{0011} & =\frac{1}{A}\Big[\frac{x^2}{r^4}(A^2r^2-B^2)-A^2\Big]-\frac{AB^2}{r^2},\\
\Upsilon^{0012} & =\frac{xy}{Ar^4}(A^2r^2-B^2),\\
\Upsilon^{0013} & =\frac{xz}{Ar^4}(A^2r^2-B^2),\\
\Upsilon^{0022} & = \frac{1}{A}\Big[\frac{y^2}{r^4}(A^2r^2-B^2)-A^2\Big]-\frac{AB^2}{r^2},\\
\Upsilon^{0023} & =\frac{yz}{Ar^4}(A^2r^2-B^2),\\
\Upsilon^{0033} & =
\frac{1}{A}\Big[\frac{z^2}{r^4}(A^2r^2-B^2)-A^2\Big]-\frac{AB^2}{r^2}.
\end{array}
\end{equation}
Using these components in (\ref{P1}), we get the following energy
and momentum density components
\begin{equation}
\begin{array}{ccc}
 \Omega^{00}& =\frac{A}{8\pi r^4}\big[r^2-B^2\big],\\
 \Omega^{10}& =-\frac{x}{8\pi r^4}\big[r^2 \dot{A} +B(\dot{A} B +
 2A\dot{B})\big],\\
\Omega^{20}& =-\frac{y}{8\pi r^4}\big[r^2 \dot{A} +B(\dot{A} B +
 2A\dot{B})\big],\\
 \Omega^{30}& =-\frac{z}{8\pi r^4}\big[r^2 \dot{A} +B(\dot{A} B +
 2A\dot{B})\big].
\end{array}
\end{equation}
Using equations (\ref{P}) in equation (\ref{emP}), we obtain the
energy and momentum distributions
\begin{equation}
E_{P}=\frac{A}{2r}(r^2+B^2)
\end{equation}
\begin{equation}
P^1=P^{2}=P^{3}=0
\end{equation}
The above results do not agree with the results obtained before,
using Einstein, Bergmann-Thomson and Landau and Lifshitz
energy-momentum complexes. A signature switch should  affect about
everything including energy distribution.
\section*{\bf{Discussion}}
We investigated the energy  and momentum (due to matter plus
fields including gravity) distribution of the Kantowski and Sachs
space-time  using  the Einstein, Bergmann-Thomson, Landau-Lifshitz
and Papapetrou energy-momentum complexes. We found that the
quantities of energy and momentum densities as well as energy
distribution are well-defined and well-behaved. For the space-time
under consideration, we found that the energy-momentum complexes
of Einstein and Bergmann-Thomson give the same results, while
Landau-Lifshitz and Papapetrou  do not give the same results and
not agree with the aforementioned complexes. We have shown that a
signature switch affects about every thing (by changing the sign
of the values of energy and momentum densities as well as energy
distribution) including energy distribution. These changes occur
in the case of Einstein and Papapetrou prescriptions but not in
Bergmann-Thomson and Landau-Lifshitz prescriptions.

\end{document}